\pdfoutput=1
\documentclass[12pt]{article}

\font\grande=cmr9.5 scaled \magstep4
\font\medio=cmr9.5 scaled \magstep2
\outer\def\beginsection#1\par{\medbreak\bigskip
      \message{#1}\leftline{\bf#1}\nobreak\medskip
\vskip-\parskip
      \noindent}
\usepackage{graphicx} 
\begin{document}
\bibliographystyle {unsrt}

\titlepage
\vspace*{1.5cm}
\begin{center}
{\grande Tensor to scalar ratio from single field magnetogenesis}\\
\vspace{15mm}
 Massimo Giovannini 
 \footnote{Electronic address: massimo.giovannini@cern.ch} \\
\vspace{1cm}
{{\sl Department of Physics, 
Theory Division, CERN, 1211 Geneva 23, Switzerland }}\\
\vspace{0.5cm}
{{\sl INFN, Section of Milan-Bicocca, 20126 Milan, Italy}}

\vspace*{1cm}
\end{center}

\centerline{\medio  Abstract}
\vspace{5mm}
The tensor to scalar ratio is affected by the evolution of the large-scale gauge fields potentially amplified during 
an inflationary stage of expansion. After deriving the exact evolution equations for the scalar 
and tensor modes of the geometry in the presence of dynamical gauge fields, 
it is shown that the tensor to scalar ratio is bounded from below by the dominance 
of the adiabatic contribution and  it cannot be smaller than one thousands 
whenever the magnetogenesis is driven by a single inflaton field.  
\vskip 0.5cm

\nonumber
\noindent

\vspace{5mm}

\vfill
\newpage
By cross-correlating the temperature and the polarization anisotropies of the Cosmic Microwave Background, the
 WMAP experiment \cite{zero,one} discovered that the initial conditions of the Einstein-Boltzmann hierarchy 
are predominantly adiabatic and Gaussian. While this conclusion is compatible 
with an inflationary origin of the large-scale curvature inhomogeneities, 
the tensor fluctuations should also produce a $B$-mode polarization
which has not been observed by the Planck experiment \cite{two}. The contribution of the tensor modes to the large-scale inhomogeneities 
is customarily parametrized in terms of the tensor to scalar ratio defined as $r_{T} = {\mathcal A}_{T}/{\mathcal A}_{{\mathcal R}}$
where ${\mathcal A}_{{\mathcal R}} = {\mathcal P}_{{\mathcal R}}(k_{p}) $ and 
${\mathcal A}_{T} = {\mathcal P}_{T}(k_{p})$ denote the amplitudes of the scalar and tensor power spectra 
 at the conventional pivot wavenumber $k_{p}= 0.002\,\, \mathrm{Mpc}^{-1}$.
According to the current data $r_{T} < 0.11$ \cite{two}. Moreover, in the case of conventional inflationary models, the tensor spectral index $n_{T}$ and the slow roll parameter\footnote{In the present discussion $H$ is the Hubble rate and the overdot denotes the cosmic time derivative.}  $\epsilon= - \dot{H}/H^2$ 
are both related by the so-called consistency relations stipulating that 
$r_{T} = 16\epsilon = - 8 n_{T}$. 

The gist of the present argument can be summarized as follows. Magnetogenesis scenarios based on the evolution 
of a single scalar field coupled to the kinetic term of the gauge fields \cite{three,four,five,six} affect the tensor and the scalar 
modes of the geometry \cite{seven} and hence modify the value of $r_{T}$ which can be reliably computed in rather 
general terms by considering the following scalar-vector-tensor action:
\begin{equation}
S= \int d^{4} x \sqrt{-g} \biggl[ - \frac{\overline{M}_{P}^2R}{2} + \frac{1}{2} g^{\alpha\beta} \partial_{\alpha} \varphi \partial_\beta \varphi - V(\varphi) 
- \frac{\lambda(\varphi)}{16 \pi} Y^{\alpha\beta} Y_{\alpha\beta}\biggr],
\label{zeroth1}
\end{equation}
where $\varphi$ is the scalar field driving the background geometry, $V(\varphi)$ is the associated potential  and $\lambda(\varphi)$ 
parametrizes the coupling of the gauge kinetic term to $\varphi$. In the case of conventional inflationary scenarios $\varphi$ 
coincides with the inflaton, however the evolution equations of the scalar and tensor modes can be derived 
without any reference to the inflationary dynamics. Note that in Eq. (\ref{zeroth1}) $8 \pi G = 1/\overline{M}_{P}^2$ while $R$ and $g$ are, respectively, the Ricci scalar and
the determinant of the four-dimensional metric $g_{\mu\nu}$. 
We shall be working in a conformally flat background metric of Friedmann-Robertson-Walker type 
denoted by $\overline{g}_{\mu\nu} = a^2(\tau) \eta_{\mu\nu}$
where $\eta_{\mu\nu}$ is the Minkowski metric with signature $(+,\,-,\,-,\, -)$. In this case, the components of the Abelian field 
strength are $Y^{i0} = e^{i}/a^2$ and $Y^{ij} = - \epsilon^{ijk} b_{k}/a^2$ while the comoving electric and magnetic fields will be denoted, respectively, by 
$\vec{E} = \sqrt{\lambda} \,a^2\, \vec{e}$ and $\vec{B}= \sqrt{\lambda} \,a^2 \,\vec{b}$; their evolution is given by: 
\begin{equation}
 \vec{E}^{\prime} +  {\mathcal F} \vec{E} = \vec{\nabla}\times \vec{B}, \qquad \vec{B}^{\prime} - {\mathcal F} \vec{B} =  - \vec{\nabla}\times \vec{E},
\label{first}
\end{equation}
where the prime indicates a derivation with respect to the conformal time coordinate and ${\mathcal F} = (\sqrt{\lambda}^{\,\prime}/\sqrt{\lambda})$ controls the 
rate of variation of the electric and magnetic fields. Equations (\ref{first}) 
are invariant under the duality transformations \cite{twoa} $\vec{E} \to -\vec{B}$, $\vec{B} \to \vec{E}$ and ${\mathcal F} \to - {\mathcal F}$. This 
observation will be relevant especially in connection with the evolution of the Poynting vector.

The tensor fluctuation of the geometry is $\delta_{t} g_{ij} = - a^2 h_{ij}$ where $h_{ij}$ is transverse and traceless (i.e. 
$\partial_{i} h^{ij} = h_{i}^{i}=0$).  In the presence of large-scale gauge fields the evolution of $h_{ij}$ is affected by the anisotropic 
stress of the gauge fields: 
\begin{equation}
h_{ij}^{\prime\prime} + 2 {\mathcal H} h_{ij}^{\prime} - \nabla^2 h_{ij} = - \frac{2 a^2}{\overline{M}_{P}^2} \biggl( \Pi_{ij}^{(E)} + 
 \Pi_{ij}^{(B)}\biggr),
 \label{third}
\end{equation} 
where, as usual, ${\mathcal H} = a^{\prime}/a = a H$ while the electric and the magnetic anisotropic stresses are defined as:
\begin{eqnarray}
\Pi_{ij}^{(E)} = \frac{1}{4 \pi a^4} \biggl[ E_{i} E_{j} - \frac{E^2}{3} \delta_{ij} \biggr], \qquad \Pi_{ij}^{(B)} = \frac{1}{4 \pi a^4} \biggl[ B_{i} B_{j} - \frac{B^2}{3} \delta_{ij} \biggr].
\label{thirda}
\end{eqnarray}
Equations (\ref{third})--(\ref{thirda}) are explicitly invariant under infinitesimal diffeomorphisms and under 
Abelian gauge transformations. 

The momentum constraint (following from the $(0i)$ components of the perturbed Einstein equations)
couples together the scalar fluctuations of the metric, the inhomogeneities of $\varphi$ and the Poynting vector. 
Consequently to reach a decoupled expression analog to Eqs. (\ref{third}) and (\ref{thirda}) it is useful to introduce an auxiliary variable $\overline{\Delta}_{{\mathcal R}}$ defined as \cite{seven}
\begin{equation}
\overline{\Delta}_{{\mathcal R}} = \Delta_{{\mathcal R}} - \frac{{\mathcal H} a^2}{\varphi^{\prime\,2}} P, \qquad P = \frac{\vec{\nabla}\cdot( \vec{E}\times \vec{B})}{4 \pi a^4},
\label{fourth}
\end{equation}
where $\Delta_{{\mathcal R}}$ is the Laplacian of the curvature perturbations on comoving orthogonal hypersurfaces (i.e. $\Delta_{{\mathcal R}} = \nabla^2 {\mathcal R}$) 
and $P$ is the three-divergence of the Poynting vector. The equation obeyed by $\overline{\Delta}_{{\mathcal R}}$ is given by: 
\begin{equation}
\overline{\Delta}_{{\mathcal R}}^{\prime\prime} + 2 \frac{z^{\prime}}{z} \overline{\Delta}_{{\mathcal R}}^{\prime} - \nabla^2 \overline{\Delta}_{{\mathcal R}}= {\mathcal S}, \qquad z= \frac{a \varphi^{\prime}}{{\mathcal H}}.
\label{fourtha} 
\end{equation}
The source term ${\mathcal S}$ does not only depend on $P$ but also on the fluctuations of the electric 
and of the magnetic energy density denoted, respectively, by  $\delta \rho_{E}= E^2/(8 \pi a^4)$ and 
$\delta \rho_{B}= B^2/(8 \pi a^4)$; more specifically ${\mathcal S}$ can be written as
\begin{equation}
{\mathcal S} = \frac{a^2}{2 \overline{M}_{P}^2}\biggl[ P^{\prime} - \biggl( 2 \frac{{\mathcal H}^{\prime}}{{\mathcal H}} + 2 \frac{a^2}{\varphi^{\prime}} V_{,\, \varphi}\biggr) P+ 
\nabla^2 (\delta \rho_{B} + \delta\rho_{E}) \biggr] + \frac{ 2 a^2 {\mathcal H} {\mathcal F} }{\varphi^{\prime\, 2}}\nabla^2(\delta\rho_{B} - \delta\rho_{E} ),
\label{fourthb}
\end{equation}
where $V_{,\, \varphi} \equiv \partial V/\partial\varphi$. Equations (\ref{fourtha})--(\ref{fourthb}) are explicitly invariant under infinitesimal diffeomorphisms and under 
Abelian gauge transformations, exactly as Eqs. (\ref{third})--(\ref{thirda}).
The actual values of $\overline{\Delta}_{{\mathcal R}}$ (or $\Delta_{{\mathcal R}}$) are the same 
in any coordinate systems but their explicit expressions do change 
from one coordinate system to the other. In the uniform curvature gauge \cite{eight} $\overline{\Delta}_{{\mathcal R}}$ coincides with the evolution of the scalar field fluctuation. Even if this is probably the most convenient 
gauge for a swift derivation of Eqs. (\ref{fourtha}) and (\ref{fourthb}), the same result can be obtained in any gauge and, in particular, in the 
longitudinal and synchronous gauges. For a closely related derivation see, in particular, the last two papers of Ref. \cite{seven}.
 
Equation (\ref{fourth}) stipulates that whenever the Poynting vector is either absent or negligible the expression of $\overline{\Delta}_{{\mathcal R}}$  coincides 
with the Laplacian of the curvature perturbations on comoving orthogonal hypersurfaces either exactly or approximately.  
This observation can be used to simplify the form of  the source term ${\mathcal S}$ appearing in Eq. (\ref{fourthb}). 
Indeed, the conservation of the total energy-momentum tensor of the gauge fields implies that the three-divergence of the Poynting vector evolves according to 
\begin{equation}
P^{\prime} + 4 {\mathcal H} P  = \nabla^2[\Pi_{\mathrm{B}} + \Pi_{\mathrm{E}} - (\delta p_{\mathrm{B}} + \delta p_{\mathrm{E}})],
\label{fourthc}
\end{equation}
where  $\delta p_{B}= \delta\rho_{B}/3$ and $\delta p_{E} = \delta\rho_{E}/3$; furthermore the following standard notations
\begin{equation}
\nabla^2 \Pi_{B}(\vec{x},\tau) = \partial_{i} \partial_{j} \Pi^{ij}_{(B)}(\vec{x},\tau), \qquad \nabla^2 \Pi_{E}(\vec{x},\tau) = \partial_{i} \partial_{j} \Pi^{ij}_{(E)}(\vec{x},\tau)
\label{fourthe}
\end{equation}
have been introduced. As already suggested, the duality symmetry of Eq. (\ref{first}) implies that the three-divergence of the Poynting vector can only be suppressed in an expanding Universe:  when the magnetic components 
are amplified the electric fields are suppressed at the same rate; vice versa when the electric fields are amplified the magnetic contribution 
is suppressed at the same rate. This is why, according to Eq. (\ref{fourthc}), $P$ (which is the three-divergence of the vector product of $\vec{E}$ and $\vec{B}$) 
can only decrease as a consequence of the expansion of the Universe. 

Therefore, over sufficiently large-scales 
(where the Laplacians at the right-hand side of Eq. (\ref{fourthc}) can be neglected), the evolution of $P$ obeys 
$P^{\prime} + 4 {\mathcal H} P=0$ implying a sharp exponential suppression of $P$ all along the conventional inflationary evolution. Thanks to this 
occurrence, up to corrections ${\mathcal O}(P)$, the evolution equations obeyed by $\delta\rho_{E}$ and $\delta\rho_{B}$ can be effectively decoupled:
\begin{equation}
 \delta \rho_{\mathrm{B}}^{\prime} + 4 {\mathcal H} \delta \rho_{\mathrm{B}} = 2 {\mathcal F} \delta\rho_{\mathrm{B}}  + {\mathcal O}(P),\qquad 
 \delta \rho_{\mathrm{E}}^{\prime} + 4 {\mathcal H} \delta \rho_{\mathrm{E}} = - 2 {\mathcal F}  \delta\rho_{\mathrm{E}}  + {\mathcal O}(P).
\label{fourthg}
\end{equation}
Inserting now Eqs. (\ref{fourthe}) and (\ref{fourthg}) into Eq. (\ref{fourthb}), a simpler expression of the source function ${\mathcal S}$ can be obtained:
\begin{equation}
{\mathcal S} =  \frac{a^2}{3\overline{M}_{P}^2}\biggl[ \nabla^2 (\delta \rho_{B} + \delta\rho_{E}) + \nabla^2(\Pi_{B} + \Pi_{E})  + 2 \biggl(\frac{z}{a}\biggr)^{\prime} \biggl(\frac{a}{z} \biggr) P\biggr] + \frac{2 a^2 {\mathcal H} {\mathcal F}}{\varphi^{\prime\, 2}}\nabla^2(\delta \rho_{B} - \delta\rho_{E}).
\label{fourthf}
\end{equation}
While the results of  Eq. (\ref{fourthf}) only assume that the background is expanding,  the expression of ${\mathcal S}$
can be further simplified by taking into account of the slow-roll dynamics. 

Equations (\ref{third}) and (\ref{fourtha}) can be solved in the long wavelength 
limit. The large-scale tensor and scalar power spectra will then be determined and from their quotient we shall deduce the wanted expression 
of the tensor to scalar ratio $r_{T}$.  The solution of Eq. (\ref{third}) for typical length scales larger than the Hubble radius 
at the corresponding epoch is  given by the sum of the adiabatic\footnote{Even if the adiabaticity condition refers not to the tensor modes (but rather to the scalar ones), we shall just use this terminology to distinguish the conventional large-scale solution from the one induced by the gauge fields.} and of the gauge contributions, i.e.
\begin{equation}
h_{ij}(\vec{x},\tau) = h^{(ad)}_{ij} + h_{ij}^{(B)} + h_{ij}^{(E)},
\label{fifth}
\end{equation}
where $h^{(ad)}_{ij}$ denotes the conventional large-scale solution of the corresponding homogeneous equation 
while the terms induced by the magnetic and electric components have the same form and can be written, in a unified notation, as:
\begin{equation} 
h_{ij}^{(X)}(\vec{x},\tau) = - \frac{2}{\overline{M}_{P}^2} \int_{\tau_{ex}}^{\tau} \frac{d\tau^{\prime\prime}}{a^2(\tau^{\prime\prime})}  \int_{\tau_{ex}}^{\tau^{\prime\prime}} \,
a^{4}(\tau^{\prime}) \, \Pi_{ij}^{(X)}(\vec{x},\tau^{\prime})\, d\tau^{\prime}.
\label{sixth}
\end{equation}
In Eq. (\ref{sixth}) the superscript is given by $X=E,\,B$ and corresponds either to the magnetic or to the electric anisotropic stress. Furthermore $\tau_{ex}$ denotes the exit time 
of a given length-scale from the Hubble radius: even if $\tau_{ex}$ has a precise meaning only in Fourier space, it can also 
be employed in real space with the aim of separating the large-scale from the small-scale solutions. 
Because of the duality symmetry of Eq. (\ref{first}) and thanks to the suppression of the Poynting vector 
(see Eq. (\ref{fourthg}))  only one of the two gauge contributions appearing in Eq. (\ref{fifth})
will be dominant for a given set of initial conditions: if the magnetic contribution increases then the 
electric contribution will decrease and vice versa. Assuming, for the sake of concreteness, that the magnetic contribution 
increases,  the electric contribution is suppressed at the same rate of the magnetic one and the dominant gauge contribution 
entering Eq. (\ref{fifth}) is
\begin{equation}
h_{ij}^{(B)}(\vec{x}, a) = - \frac{2}{g_{B} ( g_{B} + 3)}  \frac{\Pi_{ij}^{(B)}(\vec{x},a)}{H_{ex}^2\overline{M}_{P}^2}, \qquad \Pi_{ij}^{(B)}(\vec{x},a) = 
\overline{\Pi}^{(B)}(\vec{x},a_{ex}) \biggl(\frac{a}{a_{ex}}\biggr)^{g_B},
\label{seventh}
\end{equation}
where the conformal time coordinate can be traded for the scale factor in the various integrals while $g_{B}$ and $f$ are defined as: 
\begin{equation}
g_{B}= [2 f ( 1 + \epsilon) - 4 - 3 \epsilon], \qquad \int {\mathcal F} \frac{d a}{{\mathcal H} a} = f \int \frac{d a}{ a}.
\label{eight}
\end{equation}
In Eq. (\ref{eight}) $\epsilon$ denotes, as usual, the slow-roll parameter while $f$ measures, in practice,  the average growth rate ${\mathcal F}$ in units of ${\mathcal H}$. 
In the limit $\epsilon\to 0$ we have that $g_{B} = 2 f - 4$ implying that Eq. (\ref{seventh}) is singular whenever $f=2$. 
In this case the growth rate equals exactly the suppression of the energy density due 
to the expansion of the Universe. This divergence, however, only occurs in the case of the pure de Sitter dynamics (i.e.  $\epsilon \to 0$) where, strictly speaking, the 
scalar modes are absent. Moreover, if the calculation is performed, from very the beginning, for $f=2$ and $\epsilon=0$ the potential divergence is replaced by a logarithmic enhancement of the type $\ln{(a/a_{ex})}$. In spite of this possibility, since the pure de Sitter dynamics is unrealistic the slow-roll corrections must be correctly taken into account when repeatedly integrating over the conformal time coordinate. Thus, when the slow-roll corrections are included, in the limit $f \to 2$ the purported divergence disappears but $h_{ij}^{(B)}$ is enhanced by a factor going as as $1/g_{B} \to 1/\epsilon$.

Moving now to the solution of the scalar modes, we can notice that
all the terms inside the square bracket of Eq. (\ref{fourthf}) are subleading in comparison with the second term which is instead proportional to $1/\epsilon$ and hence dominant in the slow-roll regime. This statement can be easily demonstrated by appreciating that the contribution multiplying $P$ is given by:  
\begin{equation}
2 \biggl(\frac{z}{a}\biggr)^{\prime} \biggl(\frac{a}{z} \biggr) P = 2 (a H) [ 1 - \eta - \epsilon]\, P, \qquad \eta = \frac{\ddot{\varphi}}{H \dot{\varphi}}.
\label{nine}
\end{equation}
Since $\eta$ and $\epsilon$ are both negligible during the slow-roll regime, the contribution of Eq. (\ref{nine}) is simply of order $P$ and hence negligible in comparison with the others Laplacians appearing inside the square bracket of Eq. (\ref{fourthf}). Rewriting the last term at the right hand side of Eq. (\ref{fourthf}) in terms of $\epsilon$ the following inequality can be easily verified:
\begin{equation}
\frac{a^2}{3\overline{M}_{P}^2}\biggl[ \nabla^2 (\delta \rho_{B} + \delta\rho_{E}) + \nabla^2(\Pi_{B} + \Pi_{E}) \biggr] \ll \frac{a^2}{\epsilon\,\overline{M}_{P}^2} 
\biggl(\frac{{\mathcal F}}{{\mathcal H}}\biggr) \nabla^2(\delta \rho_{B} - \delta \rho_{E}). 
\label{ten}
\end{equation}
Since the definition of $\overline{\Delta}_{{\mathcal R}}$ given in Eq. (\ref{fourth}) contains exponentially suppressed corrections which are ${\mathcal O}(P)$, 
the Laplacians can be dropped on both sides of Eq. (\ref{fourtha}) so that the evolution equation of ${\mathcal R}$ takes following simple form:
\begin{equation}
{\mathcal R}^{\prime\prime} + 2 \frac{z^{\prime}}{z} {\mathcal R}^{\prime} - \nabla^2 {\mathcal R} = \frac{a^2}{\epsilon\, \overline{M}_{P}^2 } \biggl(\frac{{\mathcal F}}{{\mathcal H}} \biggr) 
 \nabla^2 (\delta\rho_{B} - \delta \rho_{E}). 
 \label{eleven}
 \end{equation}
 Equation (\ref{eleven}) can then be solved with the same methods leading to Eqs. (\ref{fifth}), (\ref{sixth}) and (\ref{seventh}). The result of this step is given by 
 \begin{equation}
{\mathcal R}(\vec{x}, a) = {\mathcal R}^{(ad)}(\vec{x}) + \frac{f \, \delta \rho_{B}(\vec{x}, a) }{\epsilon \, g_{B} ( g_{B} + 3) \, H_{ex}^2 \overline{M}_{P}^2}, \qquad \delta \rho_{B}(\vec{x}, a) = 
\delta\rho_{B}(\vec{x}, a_{ex}) \biggl(\frac{a}{a_{ex}}\biggr)^{g_{B}},
\label{twelve}
\end{equation}
where, with the same notation of Eq. (\ref{fifth}), ${\mathcal R}^{(ad)}$ denotes the genuine adiabatic contribution. In Eq. (\ref{twelve}) (as in Eq. (\ref{seventh})) the magnetic initial conditions have been assumed are assumed so that the electric contribution is eventually negligible. In the case of electric initial conditions the magnetic contribution will be instead negligible.
 
The power spectra of the scalar and tensor modes of the geometry can now be computed from Eqs. (\ref{fifth}), (\ref{seventh}) and (\ref{twelve}). 
Within the present conventions they are defined as\footnote{Note that  ${\mathcal S}_{ijmn}(\hat{k}) = [p_{m i}(\hat{k}) p_{n j}(\hat{k}) + p_{m j}(\hat{k}) p_{n i}(\hat{k}) - p_{i j}(\hat{k}) p_{m n}(\hat{k})]/4$ and $p_{ij}(\hat{k}) = (\delta_{ij} - \hat{k}_{i} \hat{k}_{j})$ denotes the standard traceless projector.} :
\begin{eqnarray}
&&\langle {\mathcal R}(\vec{k}, \tau) {\mathcal R}(\vec{p},\tau) \rangle = \frac{2 \pi^2}{k^3}\, {\mathcal P}_{{\mathcal R}}(k,\tau) \delta^{(3)}(\vec{k} + \vec{p}),  
\label{thirteenth}\\
&&\langle h_{ij}(\vec{k},\tau) \, h_{mn}(\vec{p},\tau) \rangle = \frac{2\pi^2}{k^3} {\mathcal P}_{T}(k,\tau) \, {\mathcal S}_{ijmn}(\hat{k}) \delta^{(3)}(\vec{k} +\vec{p}).
 \label{fourteenth}
 \end{eqnarray}
Since in the single-field case  the magnetic (or electric) contributions are not correlated with the adiabatic component 
the scalar and the tensor power spectra will be the sum of two separate terms namely: 
\begin{eqnarray} 
 {\mathcal P}_{T}(k) =  {\mathcal P}_{T}^{(ad)}(k) + {\mathcal Q}_{\Pi}(k,\tau), \qquad {\mathcal P}_{{\mathcal R}}(k,\tau) = {\mathcal P}_{{\mathcal R}}^{(ad)}(k) + {\mathcal Q}_{B}(k,\tau),
\label{seventeenth}
\end{eqnarray}
where  ${\mathcal P}^{(ad)}_{T}(k)$ and ${\mathcal P}^{(ad)}_{\mathcal R}(k)$ are given by: 
\begin{equation}
{\mathcal P}^{(ad)}_{T}(k) = \frac{2}{3 \pi^2} \biggl(\frac{V}{\overline{M}_{P}^4} \biggr) \biggl(\frac{k}{k_{p}}\biggr)^{n_{T}}, \qquad {\mathcal P}^{(ad)}_{\mathcal R}(k) = \frac{1}{24 \pi^2} \biggl(\frac{V}{\epsilon\, \overline{M}_{P}^4}\biggr) \biggl(\frac{k}{k_{p}}\biggr)^{n_{{s} -1}}.
\label{eighteenth}
\end{equation}
As already mentioned, $k_{p}$ denotes the conventional pivot scale at which the tensor to scalar ratio is conventionally evaluated while $n_{s}$ and $n_{T}$ are the scalar and tensor spectral indices; in Eq.  (\ref{seventeenth}) we also have that  ${\mathcal Q}_{B}(k,\tau)$ and  ${\mathcal Q}_{\Pi}(k,\tau)$ are
the power spectra of the magnetic energy density and of the magnetic anisotropic stress:
\begin{eqnarray}
&& \langle \delta\rho_{B}(\vec{k},\tau) \,\delta\rho_{B}(\vec{p},\tau) \rangle = \frac{2\pi^2}{k^3} {\mathcal Q}_{B}(k,\tau) \, \delta^{(3)} (\vec{k} + \vec{p}),
\label{eighta}\\
&& \langle \Pi^{(B)}_{ij}(\vec{k},\tau) \,\Pi^{(B)}_{mn}(\vec{p},\tau) \rangle = \frac{2\pi^2}{q^3} {\mathcal Q}_{\Pi}(k,\tau) S_{ijmn}(\hat{k})\, \delta^{(3)} (\vec{k} + \vec{p}).
\label{eightb}
\end{eqnarray}
The power spectra ${\mathcal Q}_{B}(k,\tau)$ and $ {\mathcal Q}_{\Pi}(k,\tau)$ should now be determined in terms of the magnetic power spectrum and then 
evaluated in the large-scale limit for wavenumbers comparable with the pivot scale $k_{p}$. This 
step is algebraically lengthy but standard (see, in particular, the third paper of Ref. \cite{seven}) and the result relevant for the present  purposes can be expressed as:
\begin{eqnarray}
{\mathcal Q}_{B}(k,a) &=&  H_{ex}^8 \,{\mathcal C}_{B}(f,\epsilon)\, \biggl(\frac{a}{a_{ex}}\biggr)^{2 g_{B}(\epsilon,f)} \biggl(\frac{k}{k_{\mathrm{p}}}\biggr)^{m_{B} -1},
\label{19th}\\
{\mathcal Q}_{\Pi}(k,a) &=& H_{ex}^{8}  \,{\mathcal C}_{\Pi}(f,\epsilon)\,  \biggl(\frac{a}{a_{ex}}\biggr)^{2 g_{B}(\epsilon,f)} \,\biggl(\frac{k}{k_{\mathrm{p}}}\biggr)^{m_{\Pi} -1},
\label{20th}
\end{eqnarray}
where $m_{B}= m_{\Pi} = 9 - 4 f( 1 + \epsilon)$ and the two amplitudes are instead given by:
\begin{eqnarray}
C_{B}(f,\epsilon) &=&  \frac{2^{4 f( 1 + \epsilon)}}{384\, \pi^7} \, \, \frac{[ f (1 + \epsilon) +1] \Gamma^4[f ( 1 + \epsilon) +1/2]}{ [ 4f ( 1 + \epsilon) - 5][ 4 - 2 f ( 1 + \epsilon)]},
\label{C1}\\
C_{\Pi}(f,\epsilon) &=&\frac{2^{4 f( 1 + \epsilon)}}{17280\, \pi^7}  \,\,\frac{[ 17 -2 f (1 + \epsilon)]\, \Gamma^4[f ( 1 + \epsilon) +1/2]}{ [ 
 4f ( 1 + \epsilon) - 5][ 4 - 2 f ( 1 + \epsilon)]}.
 \label{C2}
\end{eqnarray}
In the slow-roll approximation we have that $V = 3 H_{ex}^2 \overline{M}_{P}^2$. 
Therefore Eqs. (\ref{seventeenth}), (\ref{19th}) and (\ref{20th}) 
imply that the tensor and scalar power spectra at the pivot scale are:
\begin{eqnarray}
{\mathcal P}_{T}(k_{p}) &=& \frac{2}{\pi^2} \biggl(\frac{H_{ex}}{\overline{M}_{P}}\biggr)^2 + \frac{4 C_{\Pi}}{g_{B}^2 (g_{B}+3)^2}\biggl(\frac{H_{ex}}{\overline{M}_{P}}\biggr)^4\, e^{2 N_{t} g_{B}},
\nonumber\\
{\mathcal P}_{{\mathcal R}}(k_{p}) &=& \frac{1}{8\pi^2 \epsilon} \biggl(\frac{H_{ex}}{\overline{M}_{P}}\biggr)^2 + \frac{f^2 C_{B}}{g_{B}^2 (g_{B}+3)^2}\biggl(\frac{H_{ex}}{\overline{M}_{P}}\biggr)^4\, e^{2 N_{t} g_{B}},
\label{21th}
\end{eqnarray}
where the total number of efolds $N_{t}$ has been introduced.  If we now choose the pivot scale $k_{p} =0.002 \, \mathrm{Mpc}^{-1}$, $(H_{ex}/\overline{M}_{P})$ can be written in terms of the normalization of the temperature and polarization anisotropies ${\mathcal A}_{{\mathcal R}}$: 
\begin{equation}
\biggl(\frac{H_{ex}}{\overline{M}_{P}}\biggr)^2 = 8 \pi^2 \epsilon {\mathcal A}_{{\mathcal R}}, \qquad {\mathcal A}_{{\mathcal R}} = 2.41 \times 10^{-9}.
\label{21ath}
\end{equation}

Taking now the ratio of the total spectra of Eq. (\ref{21th}) and recalling the notation of Eq. (\ref{21ath}) the tensor to scalar ratio $r_{T}$
can be finally written as:
\begin{equation}
r_{T}(k_{p}) = 16 \epsilon \frac{1 + T_{\Pi}(\epsilon,f) e^{2 N_{t} g_{B}}}{1 + T_{B}(\epsilon,f)  e^{2 N_{t} g_{B}}},
\label{22th}
\end{equation}
where 
\begin{equation}
T_{\Pi}(f,\epsilon) = \frac{64 \pi^4 \epsilon {\mathcal A}_{{\mathcal R}}}{g_{B}^2 (g_{B} + 3)^2}C_{\Pi}(f,\epsilon) , \qquad T_{B}(f,\epsilon) = \frac{64 \pi^4 {\mathcal A}_{{\mathcal R}}}{g_{B}^2 (g_{B} + 3)^3} C_{B}(f, \epsilon).
\label{23th}
\end{equation}
If we now apply the simplest strategy we can consider a potential variation 
of $N_{t}$ between $50$ and $100$ while $\epsilon$ varies, for instance, between $10^{-6}$ and $0.1$. It is 
easy to see numerically that in this range, as previously suggested \cite{seven},  $f$ cannot exceed $2.2$. If the magnetic fields are to be 
amplified, the physical range for $f$ must be around $2$. To make the argument analytically more transparent consider specifically 
the case $f =2$; $T_{B}(2, \epsilon)$ and $T_{\Pi}(2, \epsilon)$ are then in a  simple relation
\begin{equation}
T_{\Pi}(2,\epsilon) = \frac{\epsilon ( 15 - 2 \epsilon)}{45 ( 3 + 2 \epsilon)} T_{B}(2,\epsilon), \qquad T_{B}(2,\epsilon) \simeq - \frac{3\, {\mathcal A}_{{\mathcal R}}}{8\pi \epsilon^3} e^{ 2 \epsilon N_{t}}.
\label{24th}
\end{equation}
Equation (\ref{24th}) has been 
obtained by neglecting the $\epsilon$ dependence in the Euler Gamma functions (see Eqs. (\ref{C1}) and (\ref{C2})), by keeping the exponential dependence 
on the total number of efolds and by expanding the remaining prefactor in powers of $\epsilon$. 
The result is sufficiently simple and accurate to explain why a lower bound on the tensor to scalar ratio is expected:
to be compatible with dominant adiabatic mode we should require, in Eqs. (\ref{22th}) and (\ref{24th}), that 
$T_{\Pi} < 0.1$, $T_{B}< 0.1$ and $r_{T} < 0.1$. Since these conditions are verified 
in a rather narrow slice of the parameter space (i.e. $0.001< \epsilon < 0.01$) we will also have that
$0.01 < r_{T} < 0.1$ for  $f =2$. If $f >2$ the bound on $r_{T}$ is relaxed but the total number of efolds is bounded from above. 
If, for instance, $f= 2.1$ Eq. (\ref{eight}) implies that $g_{B} = 0.2 + 1.2 \epsilon$ which explains why $N_{t}$ cannot be too large.
Already for $f = 2.06$ we have that the dominance of the adiabatic mode and the bounds on the tensor to scalar ratio imply 
$N_{t} < 56$ and $10^{-4} < r_{T} < 0.1$. 

All in all the logic developed in this investigation strongly suggests that whenever  $2< f < 2.2$ we must demand,  in a conservative perspective, that 
\begin{equation}
10^{-3} < r_{T} < 0.1, \qquad 50 < N_{t} < 75.
\label{25th}
\end{equation}
If the measured value of $r_{T}$ will turn out to be smaller than $10^{-3}$, single field magnetogenesis models
will be under pressure. The dynamical framework could still be viable when the gauge kinetic term is coupled 
to some other spectator field different form the inflaton \cite{nine}. In this case the tensor to scalar ratio may be smaller 
but an entropic mode will be generated and independently constrained by the temperature and polarization anisotropies.
Consequently, an excessively small tensor to scalar ratio (i.e. below one thousands) will preferentially pin down those 
scenarios characterized by spectator fields leading to negligible entropic contributions over large scales.

\end{document}